\documentclass{llncs}
\usepackage{bussproofs}
\usepackage{amssymb}
\usepackage{graphicx}
\usepackage{url}
\usepackage{color}
\usepackage{proof}
\usepackage{qtree}
\newcommand\tool[1]{\textsf{#1}\space}
\newcommand\vmdv{\tool{VMDV}}
\newcommand\couic[1]{}

\begin{document}

\title{\vmdv: A 3D Visualization Tool for Modeling, Demonstration, and Verification}
\author{Jian Liu\inst{1,2}\and Ying Jiang\inst{1} \and Yanyun Chen\inst{1}\and Qing Zhou\inst{1,2}}
\institute{State Key Laboratory of Computer Science\\ Institute of Software, Chinese Academy of Sciences
\and University of Chinese Academy of Sciences
\\\email{\{liujian, jy, chenyy, zqing\}@ios.ac.cn}}

\maketitle

\begin{abstract}
The output of an automated theorem prover is usually presented by using a text format,
they are often too heavy to be understood. 
In model checking setting, it would be helpful if one can observe the structure of models and the verification procedures. 
A 3D visualization tool (\textsf{VMDV}) is proposed in this paper to address these problems. 
The facility of \vmdv is illustrated by applying it to a proof systems.

 \end{abstract}

\section{Introduction}
In the field of mathematical logic, a formal proof is usually defined as a sequence of formulas,
which are either axioms or logical consequences of the preceding formulas. 
In the most natural way, a proof is generally presented as a proof tree, where
each node is labelled by a formula, and its sub-nodes are the hypothesises. 
Nowadays, one can obtain proofs from computers automatically,
thanks to the development of automated theorem provers. 
However, the outcome of a theorem prover is usually displayed in text format. This usually makes the proof tree difficult to understand (e.g., to observe the relative locations of some nodes in the same proof tree), especially when the proof tree contains a large set of nodes. The same problem also arises in the field of model checking \cite{CBJPK}, where the counterexample generated by model checkers are usually hard to read. A more readable form will be helpful to engineers who are not familiar with model checking \cite{CGMZ}. Text format is difficult to show the structures when proof trees are dynamically updated. This is very common in a proving procedure, where the nodes may be dynamically created or deleted.
Then the following questions arises that motivate our work: 
\couic{
1. How to enable the output of a theorem prover to carry huge amount of information, and at the same time be easy to understand and reason about? 2. How to observe the verification procedure when checking a proof tree\couic{property of a given complex model} in an intuitive manner?
}

\begin{enumerate}
\item How to enable the output of a theorem prover to carry huge amount of information, and at the same time be easy to understand and reason about?
\item How to observe the verification procedure when checking a proof tree\couic{property of a given complex model} in an intuitive manner?
\end{enumerate}

To answer the previous questions, 
a 3D visualization tool, called \vmdv (Visualization for Modeling, Demonstration, and Verification), 
is proposed in this paper.
\vmdv employs a 3D renderer to plot proof trees in 3D space. In this way, the original text format is easily organized and displayed. Local detail text information adheres then to the node on demand.
\vmdv is based on the 3D information visualization techniques. 
As is well known, 3D information visualization techniques take advantage of the human eyes' broad bandwidth pathway into the mind to allow users to capture large amounts of information at once. In our case, \vmdv represents the proofs by 3D trees (Fig. \ref{fig:compare_text_graph_detail}). In the following figure, the 2D format is much more convenient to reflect the structure of the proof tree than the text format. However, when the number of nodes becomes large, there would not be enough space to show the labels of each node in 2D format. 3D format, on the other hand, can use space much more efficiently than 2D format when the nodes becomes large enough, by showing clearly both the structure of the graph, and the labels of each node. In this situation, 3D format can be seen as the combination of multiple 2D spaces in a consistent manner.
\begin{figure}[!h]
\centering
\includegraphics[width=10cm]{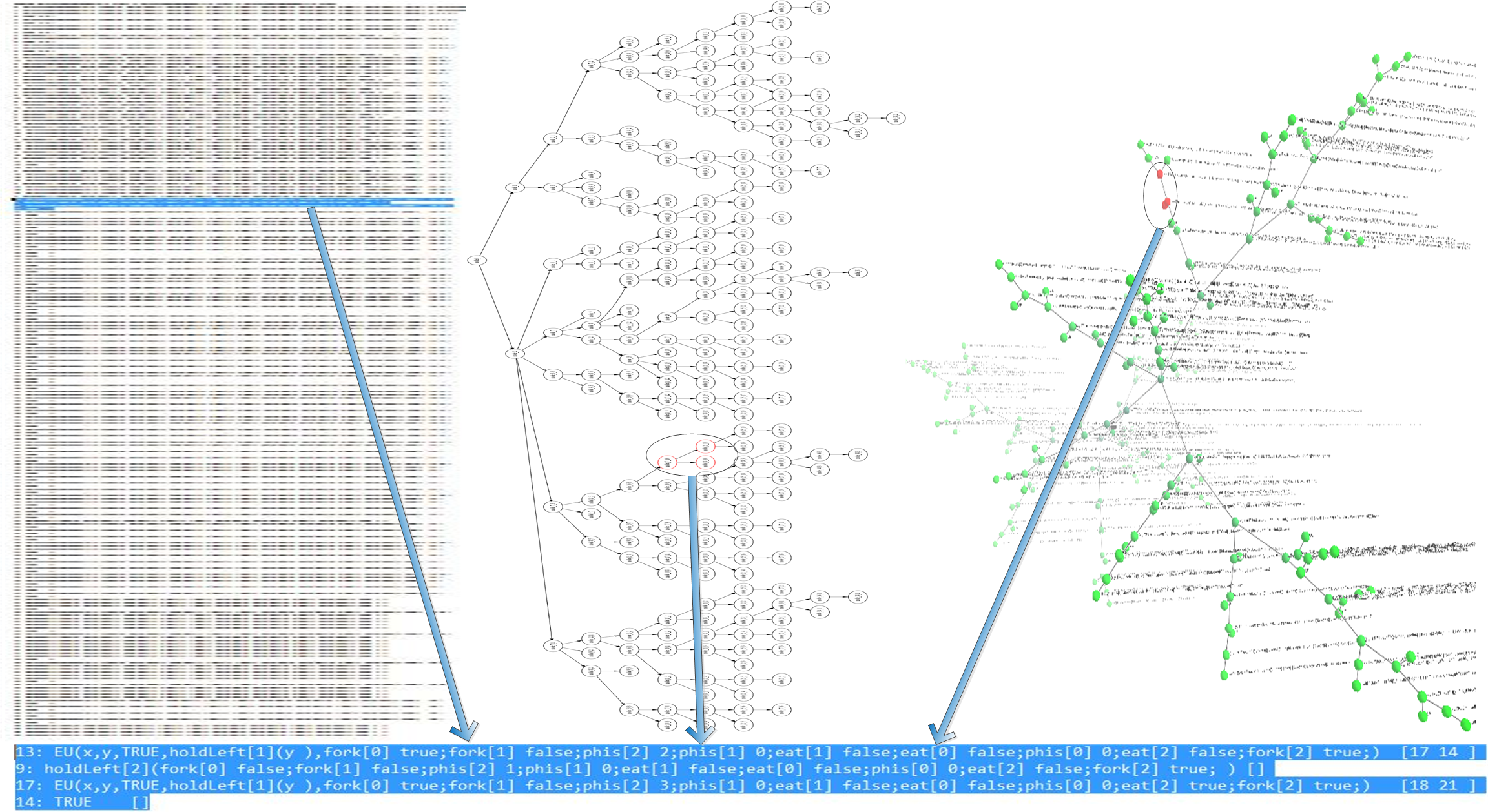}

\caption{Display of a proof tree in text format, 2D graph, and 3D graph, respectively.} 
\label{fig:compare_text_graph_detail}
\end{figure}
\vmdv allows the observation either from a global perspective or a local one. The global view shows the shape, or the topological structure while the local view shows the detailed information of interesting nodes or sub-structures 
(Fig. \ref{fig:screenshot}). 

\vmdv also enables us a variety of manners to observe or interact with the proof tree. 
The zooming and rotation operations effectively change the viewport, making the presentation of the overall structure of the proof tree clear via different angles of view (Fig. \ref{fig:prooftree_angles}); 
when the proof tree is very large and only a subtree is interesting, this subtree can be selected and get focused with other parts of the tree hidden; nodes with specific pattern can be searched and highlighted on the proof tree (Fig. \ref{fig:high_different}), etc. This allows us to find, among others, all the formulas from which one specific formula can be inferred or deduced.
\couic{
The same scenario holds in the setting of model checking (Fig. \ref{fig:state_angles}).
 }

In some proof systems, there may be some auxiliary structures emerging along with the construction of proof trees, such as Kripke models or inductively defined terms. \vmdv enables the visualization of both proof trees and auxiliary structures, and the interaction between proof trees and the auxiliary structures may help better understand the proving procedure.

As a visualization tool, \vmdv is designed and implemented as a stand-alone program, not as a part of specific proof systems. Proof systems and \vmdv communicate via TCP sockets. This facilitate the extensibility of \vmdv to different proof systems. Design such a visualization interface for existing proof systems such as Coq is our future work. 

\paragraph{\textbf{Related Work.}}As far as we know, many efforts have been made to the visualization of the output of theorem provers. 
For instance, in \cite{Farmer200939,byrnes2009visualizing,sakurai2011mikibeta,steel2005visualising}, proofs are presented in graph format instead of lines of text, and colors are used to highlight crucial parts of a proof tree; \cite{LibalRR14} proposed several criteria for the visualization of proof trees, such as distinguishing different kinds of rules, following the progress of subproofs, focusing on different aspects of the proof, etc; and \cite{bajaj2003interactive} has explored the visualization of both proof trees, and other data structures such as expressions. 
Our work is different from the existing visualization tools:
Firstly, we render the graphs in 3D format, instead of 2D format in most of the existing visualization tools. Rendering in 3D space enables the visualization of graphs with complicated structures that are usually difficult to view in 2D space. For instance, in 2D space, the graph will be confusing with hundreds of nodes, whereas in 3D space, we can demonstrate the structure of the graph with thousands of nodes clearly. 
Secondly, even though some of the existing visualization tools \cite{Farmer200939,bajaj2003interactive} use 3D libraries, their layout algorithms are rather limited to graphs with simple structures. 
Instead, we use an automatic layout algorithm which simulates a physical system where nodes repulse each other, much like magnets while edges attract the nodes that they connect, much like springs. 
This algorithm is capable of handling the layout of the proof tree smoothly, where nodes may be hidden, created, or deleted during a proving procedure.
\couic{
more complicated graphs, making the visualization of complicated graphs more comprehensible.
}

The rest of the paper is organized as follows: Section 2 presents the preliminaries of this paper, including some basic notions of information visualization, and the brief introduction of a proof system.  Section 3 introduces our visualization tool \textsf{VMDV}. Section 4 involves the applications of \vmdv to the proof system. Section 5 concerns conclusion and future works.

\section{Preliminaries}
We present the concept of information visualization, and a proof systems in sequent calculus style.
The interested reader is referred to \cite{dowek2013logical} for further details.

{
\subsection{Information Visualization}
Information visualization is the study of visual representation of abstract data that focus on the creation of approaches for humans to capture abstract information intuitively. With the visual representations of data, it is easier for humans to get a deeper understanding, and gain the essentials over the massive data-sets.

Benefit from the study of computer graphics, information presentation using visualization has been enjoying popular support.  The Open Graphics Library (\textsf{OpenGL}) provides a language independent and cross-platform application programming interface (\textsf{API}) for the rendering of three dimensional graphics. As for the hardware, the enhancing of the computational power of Graphics Processing Units (\textsf{GPU}) has made the efficient rendering of complex graphics a reality. As opposed to digital numbers which are more readable for computers, Visualization systems are more friendly to human beings for providing both concrete and abstract inspirations. For instance, plotted charts are better understandable than bare data-sets. It assists in uncovering the trends, reveal insights, or even tell stories. Information visualization provides a easier way for users to capture the abstract patterns of the massive data by applying a graphical presentation.

We applied the ideas of information visualization to the development of \textsf{VMDV}, as it provides:
\begin{itemize}
	\item The presentation of data in 3D space, where the data points are encoded as 3D solid spheres, and the structure of data are encodes as lines between spheres;
	\item Visual interaction with data, such as highlighting the search results for specific data, or controlling the progress of producing new data.
\end{itemize}
\couic{
Visual presentation and visual interaction brings much convenience in the proving procedure. As both the global and local views of the output are demonstrated at the same time that helps to understand the output.
}
}
\subsection{The System \textsf{SCTL}}
Model checking \cite{CGP01,EmersonC82,EmersonH85} and automated theorem proving
\cite{Fitting96,Loveland78} are two major pillars of formal verification methods.
The proof system \textsf{SCTL}, introduced in \cite{dowek2013logical}, is a sequent calculus for 
Computation Tree Logic (\textsf{CTL}) \cite{EmersonC82,EmersonH85},
taking a Kripke model as parameter.
\textsf{SCTL} performs verification directly on a given Kripke model from the perspective of automated theorem proving,
and produces formal proofs when the verification procedure terminates.

The syntax of \textsf{SCTL} is stipulated as follows.
The properties of a Kripke model are expressed in a language tailored for this model.
The language contains,
for each state $s$ of the model, a constant also written $s$;  
for each relation $P$ over the model, a predicate symbol, also written $P$.
Formulas are built in the usual way with the connectors $\top$, $\bot$, $\wedge$, $\vee$ and $\neg$, to which we add modalities $AX$, $EX$, $AF$, $EG$, $AR$, and $EU$. If $\phi$ is a formula, and $t$ is either a constant or a variable, then $AX_x(\phi)(t)$, $EX_x(\phi)(t)$, $AF_x(\phi)(t)$, and $EG_x(\phi)(t)$ are formulas. Like quantifiers, modalities bind the variable $x$ in $\phi$. If $\phi_1$ and $\phi_2$ are formulas and $t$ is either a constant or a variable, then $AR_{x,y}(\phi_1,\phi_2)(t)$ and $EU_{x,y}(\phi_1,\phi_2)(t)$ are formulas. These modalities bind the variable $x$ in $\phi_1$ and $y$ in $\phi_2$.

The semantics of a \textsf{SCTL} formula is defined as follows, and the proof rules for the system \textsf{SCTL} is depicted in Fig. \ref{fig:sctl_rules}.

\begin{definition}[Valid formula]
Let $\mathcal{M}$ be a model and $\phi$ be a closed formula, the set of valid formulas $\models \phi$ in the model $\mathcal{M}$ is defined by induction on $\phi$:

\begin{itemize}
        \item $\models P(s_1,...,s_n)$, if $\langle s_1,...,s_n\rangle \in P$; \  \  \  $\models \neg P(s_1,...,s_n)$, if $\langle s_1,...,s_n\rangle \notin P$,
        \item $\models \top$, \  \   \  $\models \bot$ is never the case;
        \item $\models \phi_1\wedge\phi_2$, if $\models \phi_1$ and $\models \phi_2$,
        \item $\models \phi_1\vee\phi_2$, if $\models \phi_1$ or $\models \phi_2$,
        \item $\models AX_x(\phi_1)(s)$, if for each state $s'$ in $Next(s)$, $\models (s'/x)\phi_1$,
        \item $\models EX_x(\phi_1)(s)$, if there exists a state $s'$ in $Next(s)$ such that $\models (s'/x)\phi_1$,
        \item $\models AF_x(\phi_1)(s)$, if for all infinite paths $s_0,s_1,...$ starting from $s$, there exists a natural number $i$, such that $\models (s_i/x)\phi_1$,
        \item $\models EG_x(\phi_1)(s)$, if there exists an infinite path $s_0,s_1,...$ starting from $s$, such that for all natural numbers $i$, $\models (s_i/x)\phi_1$,
        \item $\models AR_{x, y}(\phi_1,\phi_2)(s)$, if for all infinite paths $s_0,s_1,...$ starting from $s$, and for all $j$, either $\models (s_j/y)\phi_2$ or there exists an $i<j$ such that $\models (s_i/x)\phi_1$,
        \item $\models EU_{x, y}(\phi_1,\phi_2)(s)$ if there exists an infinite path $s_0,s_1,...$ starting from $s$ and a natural number $j$ such that $\models (s_j/y)\phi_2$ and for all $i<j$, $\models (s_i/x)\phi_1$.
\end{itemize}
\end{definition}

\couic{
{\color{red}{As is shown in the proof rules, for the proof of sequents with modality $AX$, $EX$, $EG$, $AR$ or $EU$, each step of unfolding of its hypothesis corresponds to the unfolding of the transition relation of the given Kripke model. That is, the Kripke model emerges along with the construction of the proof tree. The proof tree produced by \textsf{SCTL} may be too complicated or large to be comprehensible when output as text format, or even in 2D space. 
With 3D visualization tools such as \vmdv, it is more easier to understand the proof tree as it provides more insights into the overall structure and specific proof patterns. With the visualization of the proof tree as well as the Kripke model, it is straightforward to locate specific states and the formulae verified on them.}}}

\begin{figure}[t]
\noindent\framebox{\parbox{.98\textwidth}{\hspace*{-0.3cm}
$$\scriptsize
\begin{array}{@{}lll@{}}
\infer[^{\mbox{atom-\textsf{R}}}_{\langle s_1,...,s_n\rangle \in P}]{\vdash P(s_1,...,s_n)}{}
&
\hspace{5mm} \infer[^{\mbox{$\neg$-\textsf{R}}}_{\langle s_1,...,s_n\rangle \notin P}]{\vdash \neg P(s_1,...,s_n)}{}
&
\hspace{5mm} \infer[^{\mbox{$\top$-\textsf{R}}}]{\vdash \top}{}

\vspace{2mm}\\

\infer[^{\mbox{$\wedge$-\textsf{R}}}]{\vdash \phi_1\wedge\phi_2}{\vdash \phi_1 & \vdash \phi_2}
&
\hspace{5mm} \infer[^{\mbox{$\vee$-{$\mathsf{R_1}$}}}]{\vdash \phi_1\vee\phi_2}{\vdash \phi_1}
&
\hspace{5mm} \infer[^{\mbox{$\vee$-{$\mathsf{R_2}$}}}]{\vdash \phi_1\vee\phi_2}{\vdash \phi_2}
\vspace{2mm}\\

\infer[^{\mbox{$\mathbf{EX}$-\textsf{R}}}_{s'\in \textsf{Next}(s)}]{\vdash EX_x(\phi)(s)}{\vdash (s'/x)\phi}
&
\multicolumn{2}{l}{\hspace{5mm} 
\infer[^{\mbox{$\mathbf{AX}$-\textsf{R}}}_{\{s_1,...,s_n\}=\textsf{Next}(s)}]{\vdash AX_x(\phi)(s)}{\vdash (s_1/x)\phi & \ldots & \vdash (s_n/x)\phi}
}
\vspace{2mm}\\

\infer[^{\mbox{$\mathbf{AF}$-{$\mathsf{R_1}$}}}]{\Gamma \vdash AF_x(\phi)(s)}{\vdash (s/x)\phi}
&
\multicolumn{2}{l}{\hspace{5mm} 
\infer[^{\mbox{$\mathbf{AF}$-{$\mathsf{R_2}$}}}_{\{s_1,...,s_n\}=\textsf{Next}(s)}]{\Gamma \vdash AF_x(\phi)(s)}{\Gamma\vdash AF_x(\phi)(s_1) & \ldots & \Gamma\vdash AF_x(\phi)(s_n)}
}
\vspace{2mm}\\

\multicolumn{2}{l}{
  \infer[^{\mbox{$\mathbf{EG}$-\textsf{R}}}_{s' \in \textsf{Next}(s)}]{\Gamma \vdash EG_x(\phi)(s)}{\vdash (s/x)\phi & \Gamma,EG_x(\phi)(s)\vdash EG_x(\phi)(s')}
}
&
\infer[^{\mbox{$\mathbf{EG}$-\textsf{merge}}}_{EG_x(\phi)(s)\in \Gamma}]{\Gamma \vdash EG_x(\phi)(s)}{}
\vspace{2mm}\\

\multicolumn{3}{c}{
  \infer[^{\mbox{$\mathbf{AR}$-{$\mathsf{R_1}$}}}_{
{
\begin{tabular}{@{}l}
{\tiny $\{s_1,...,s_n\}=\textsf{Next}(s)$}\\
{\tiny $\Gamma' = \Gamma,AR_{x,y}(\phi_1,\phi_2)(s)$}
\end{tabular}
}
}
]{\Gamma \vdash AR_{x,y}(\phi_1,\phi_2)(s)}{\vdash (s/y)\phi_2 & \Gamma'\vdash AR_{x, y}(\phi_1,\phi_2)(s_1) ~...~ \Gamma'\vdash AR_{x, y}(\phi_1,\phi_2)(s_n)}
}
\vspace{2mm}\\

\multicolumn{3}{l}{
\infer[^{\mbox{$\mathbf{AR}$-{$\mathsf{R_2}$}}}]{\Gamma \vdash AR_{x,y}(\phi_1,\phi_2)(s)}{\vdash (s/x)\phi_1 & \vdash (s/y)\phi_2}
~~~~~~~~
\infer[^{\mbox{$\mathbf{AR}$-\textsf{merge}}}_{AR_{x,y}(\phi_1,\phi_2)(s)\in \Gamma}]{\Gamma \vdash AR_{x,y}(\phi_1,\phi_2)(s)}{}
}
\vspace{2mm}\\

\multicolumn{3}{l}{
  \infer[^{\mbox{$\mathbf{EU}$-{$\mathsf{R_1}$}}}]{\Gamma\vdash EU_{x,y}(\phi_1,\phi_2)(s)}{\vdash (s/y)\phi_2}
~~~~~~~
  \infer[^{\mbox{$\mathbf{EU}$-{$\mathsf{R_2}$}}}_{s'\in \textsf{Next}(s)}]{\Gamma\vdash EU_{x,y}(\phi_1,\phi_2)(s)}{\vdash (s/x)\phi_1 & \Gamma\vdash EU_{x,y}(\phi_1,\phi_2)(s')}
  }
\end{array}
$$

}}
\caption{\textsf{SCTL}$({\cal M})$}
\label{fig:sctl_rules}
\end{figure}

\section{\vmdv}
\subsection{Architecture}
\begin{figure}[!h]
\scriptsize
\centering
\includegraphics[width=10cm]{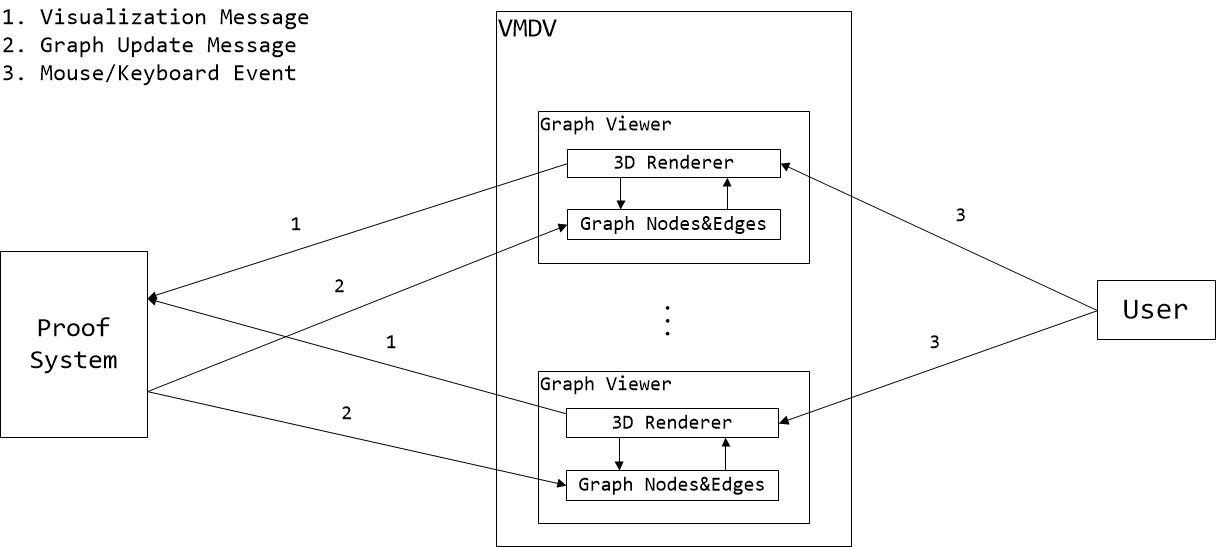}
 
\caption{The architecture of \textsf{VMDV}.}
\label{fig:architecture}
\end{figure}

 \couic{
\vmdv is designed as an independent tool, or a plug-in, for various proof systems. In our implementation, it is a standalone visualization program who communicates with the proof system via messages (Fig. \ref{fig:architecture}).
 }
 
In the design of \vmdv, the proof system does not know exactly how \vmdv renders graphs, and \vmdv does not know the working procedure of the proof system. The proof system simply sends output information to the \textsf{VMDV}, and gets the manipulation feedbacks from it. As is shown in Fig. \ref{fig:architecture}, there are two kinds of messages interchanged by a proof system and \vmdv: the Visualization Messages sent by \vmdv, and the Graph Update Messages sent by the proof system. Visualization Messages are control messages consist of requiring the related information of the current formula, such as the hypothesis, sub-formulae, or other specific information (e.g., the related state in an SCTL formula). Graph Update Messages are data messages consist of the responds to the control messages.
\vmdv serves as the interface to dynamically visualize the output of the proof system. The relationship of a \vmdv with a proof system is very similar to that of a web browser and a web server. In order to extend the applications of \vmdv to other proof systems, \vmdv is designed and implemented as a stand-alone program, not as a part of specific proof systems. Proof systems and \vmdv communicate via TCP sockets. Both control and data messages are wrapped as TCP packets.  This way, \vmdv can easily communicate with proof systems implemented in different programming languages, or run in different computers in networks.
 
Note that \vmdv allows multiple outputs to visualize multiple structures, proof trees, or, e.g., Kripke models in \textsf{SCTL} system.

\vmdv is implemented in Java\footnote{\url{https://www.oracle.com/java/index.html}} and rendered using \textsf{JOGL},\footnote{\url{http://jogamp.org/jogl/www/}} the Java binding of the \textsf{OpenGL} \textsf{API}.

\subsection{Interfaces}
\begin{figure}[h]
\centering
\includegraphics[width=6.5cm]{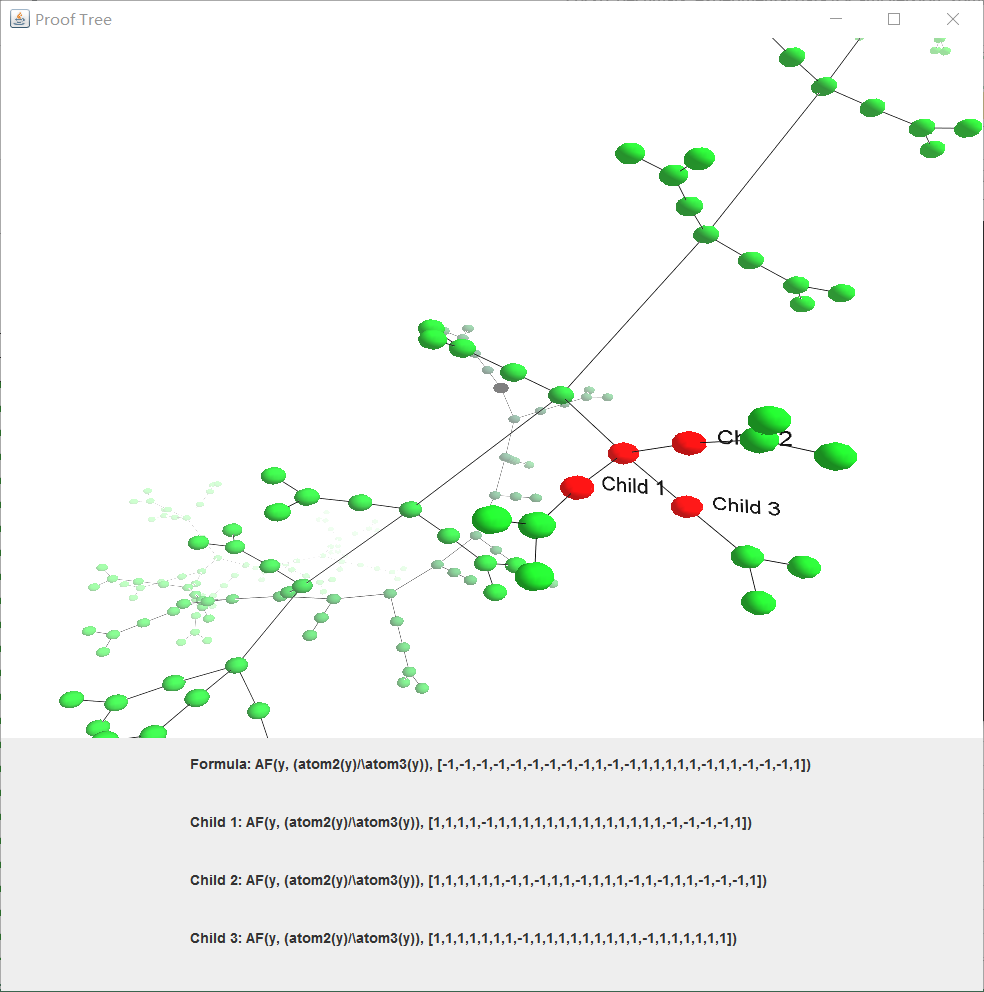}
\caption{A typical screenshot of \textsf{VMDV}.}
\label{fig:screenshot}
\end{figure}

Fig. \ref{fig:screenshot} shows a typical screenshot of \textsf{VMDV}. It consists of two panels: the main panel on the top shows the overall structure of the proof tree, and the panel at the bottom shows the details of the selected nodes.
 
Similar to other 3D visualization tools, \vmdv adopts some commonly used operations (for instance, zooming, rotation, and selection) for users to interact with 3D graphs. Furthermore, \vmdv provides mechanisms to extend its functionalities to fulfill the special requirements of different kinds of proof systems.
 
\begin{figure}[h]
\centering
\includegraphics[width=11cm]{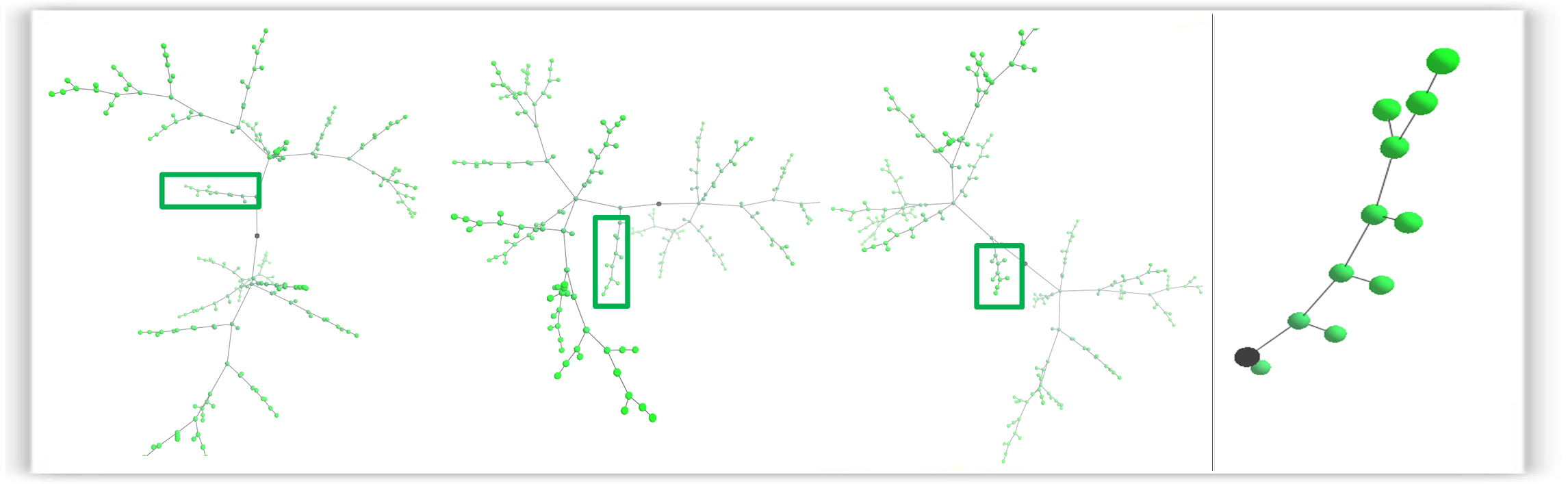}
\caption{A proof tree observed from different angles of view.}
\label{fig:prooftree_angles}
 
 \couic{
\includegraphics[width=12cm]{./imgs_backup/state_angles.png}
\caption{A Kripke model, containing 1641 nodes and 1851 edges, observed from different angles of view.}
\label{fig:state_angles}
}
\end{figure}

{\bf Zooming and rotation.}
The most obvious advantage of 3D visualization over 2D one is the capacity of observing a graph from different angles. Although there exist different ways to plot a 2D graph, it is still hard to match the 3D solution when the structure of the graph is too complicated to present in 2D space. 3D visualization techniques handle this easily by two operations, zooming and rotation: zooming in to see the details, rotating the tree to locate the sub-tree of interest, and zooming out to capture the overall shape, as are shown in Fig. \ref{fig:prooftree_angles}.
\couic{
 and Fig. \ref{fig:state_angles}.
 }

\begin{figure}[h!]
\centering
\includegraphics[width=9cm]{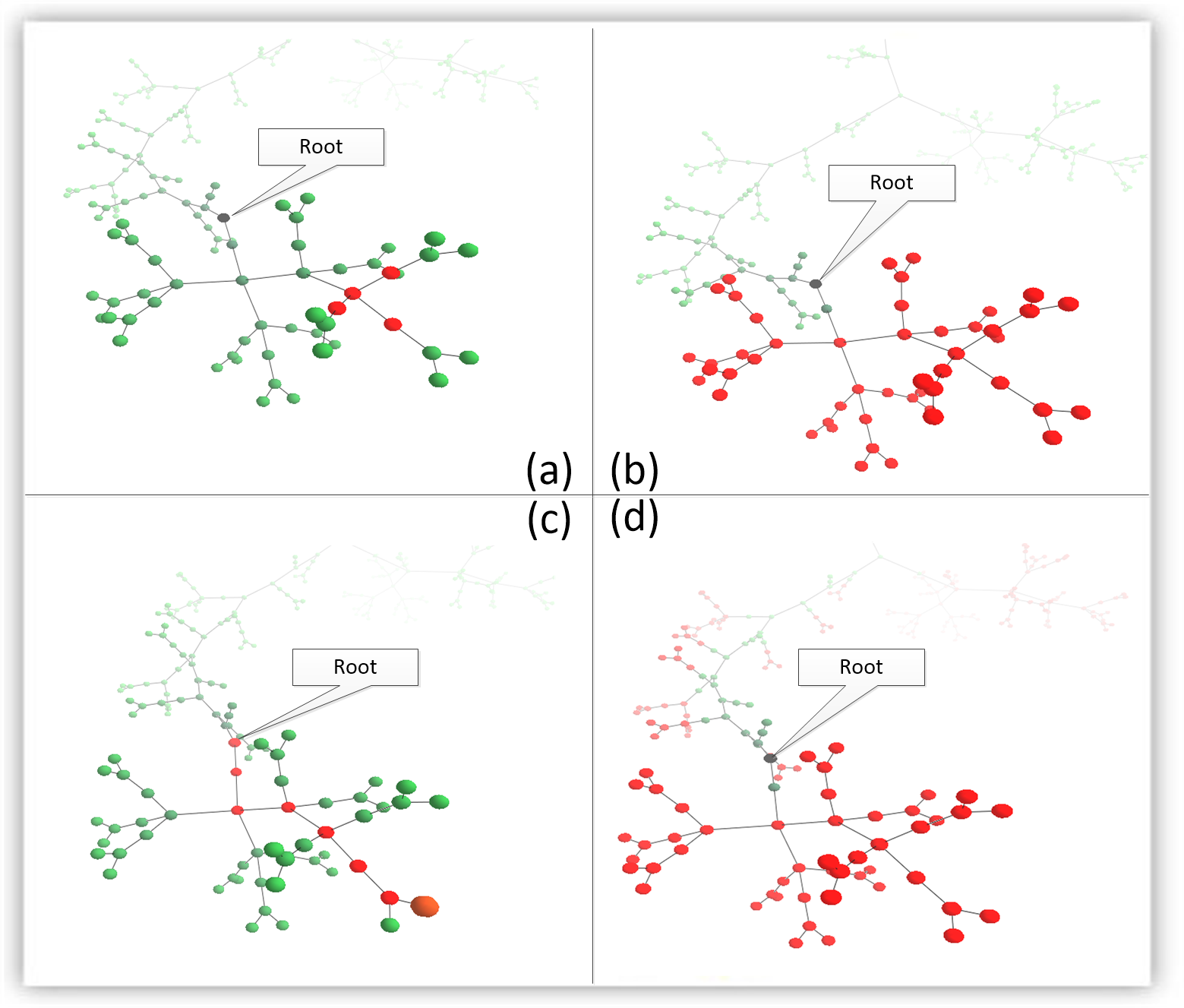}
\caption{Different kinds of highlighting: (a) selection of a single node and its children; (b) selection of a subtree; (c) highlighting ancestors of a node; (d) highlighting similar proof patterns.}
\label{fig:high_different}
\end{figure}
 
\begin{figure}[h!]
\centering
\includegraphics[width=10cm]{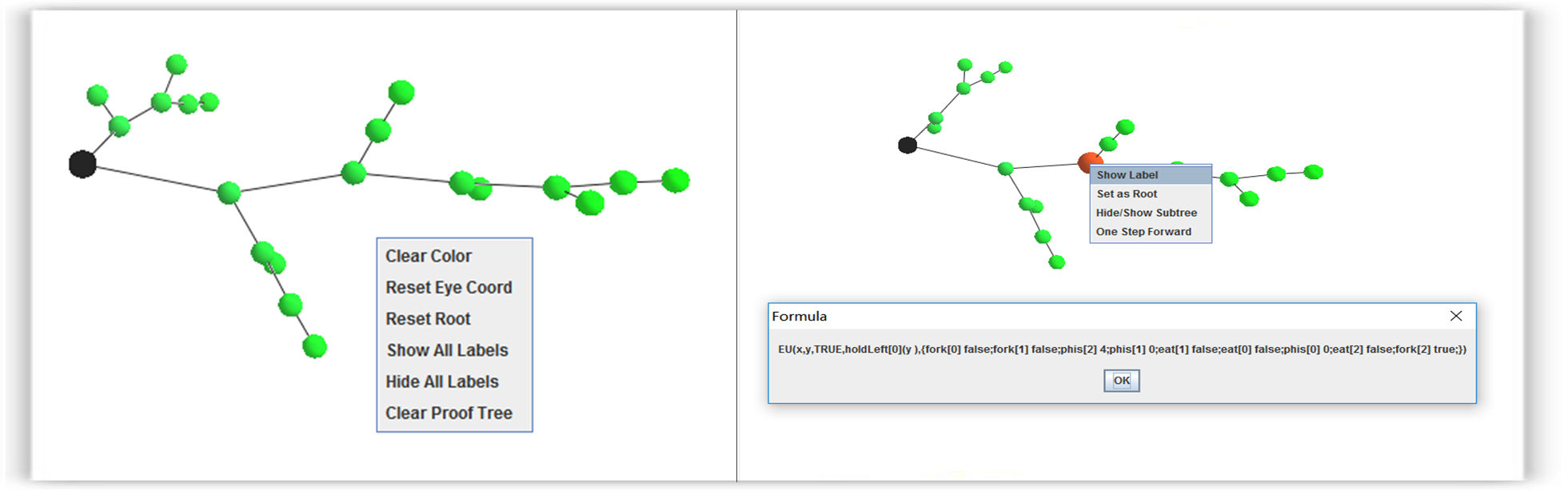}
\caption{Extensible operations.}
\label{fig:user_defined_operation}
\end{figure}
 
{\bf Highlighting.}
Sometimes some local information (for instance, specific nodes, edges, or proof patterns) is more interesting rather than the overall structure. Highlighting becomes a useful operation to show the local focused parts of the given proof tree. \vmdv enables highlighting parts of the proof tree either by manual selection using mouse clicking, or by automatic selection via formulae searching. (Fig. \ref{fig:high_different})\footnote{In \vmdv, we say that the proofs of two formulas have similar pattern if the first rules applied in the two bottom-up proofs are the same. For instance, if the proofs of two formulas in \textsf{SCTL} have similar pattern, then according to the rules in Fig. \ref{fig:sctl_rules}, both formulas must have the same modality or connector, or both are atomic formulas with the same predicate.}
 
{\bf Focusing a specific subtree.}
In many cases, interesting information may be carried out by specific subtrees. To concentrate on these subtrees, the nodes that are not related are hidden. Additional operations can then be applied only on these subtrees. The focusing of a subtree is easily done by changing the root node. A reset operation will recover the entire tree (Fig. \ref{fig:prooftree_angles}).

{\bf Extended Operations.}
In addition to the commonly used operations, some other operations that are specific to proof systems are also needed, so that \vmdv can be adapted to different proof systems. The controlling of the construction of both the proof tree and the Kripke model in the \textsf{SCTL} system is a good example. Define a new operation includes the handling of messages from \vmdv and the responds from the proof system. New operations are implemented as plug-ins that are dynamically loaded by \vmdv{}. One selection from a pop up menu triggers one such operation when right click the mouse (Fig. \ref{fig:user_defined_operation}).

\subsection{Automatic Layout}
The layout of 3D graphs is non-trivial. Our solution is modified from ForceAtlas2 \cite{jacomy2014forceatlas2}. ForceAtlas2 is a force-directed algorithm that simulates a physical system, where nodes repulse each other like magnets while edges attract the nodes they connect like springs. The forces of repulsion and attraction make the movement of nodes until the 3D graph reaches a balanced state. Since the clustering of nodes are not the main concern of \textsf{VMDV}, we set the degree of each node to be constant $1$ in our algorithm, not the number of edges that are attached to the node as ForceAtlas2 does. Same as ForceAtlas2, our algorithm is continuous and homogeneous, as opposed to the OpenOrd \cite{martin2011openord} algorithm which is not, and the algorithm of Yifan Hu \cite{yifanhu05} which is semi-continuous. The continuity and homogeneity of the algorithm make the movement of nodes to respect the feeling of a physical system, and users comprehend the 3D graph by ``playing" with it. 

\section{Applications}
In this section, we first show the applications of \vmdv in the implementation of \textsf{SCTL} (in programming language OCaml), and then illustrate how \vmdv visualize proof trees produced by existing proof system Coq \cite{bertot2013interactive}.

\subsection{The System \textsf{SCTL}}
\couic{
The particular aspects of \textsf{SCTL} implementation are: 
(1) It performs verification automatically and directly over any given Kripke model. 
(2) It generates a counterexample when the verification of the given property fails
and permits to give a certificate (proof tree) for the property when it succeeds. 
(3) It performs verification in a continuation-passing style and a double on-the-fly style, thanks
to the syntax and inference rules of \textsf{SCTL}.
}
With the application of \textsf{VMDV}, we can show both proof trees and Kripke models in 3D format. 
We can also visualize, in 3D format, the verification procedure, 
revealing gradually the relation between a proof tree and the corresponding states of the Kripke model under consideration. 
Although the Kripke model in realistic cases may be very large, thanks to the on-the-fly style of proof search in \textsf{SCTL}, we can only show the states that need to be explored, which may be a small part of the whole model.
The application of \vmdv to the proof system \textsf{SCTL} is illustrated in the following small example.

\begin{example}[The River Crossing Puzzle.]
\label{expl:river}
A farmer is trying to transport a wolf, a sheep, and a cabbage from one side of a river to another, however, he can only carry at most one item each time. During the transportation, the sheep cannot be left alone with the wolf, nor the cabbage can be left alone with the sheep.
The question is how can the farmer get across the river by bringing the wolf, the goat, and the cabbage.
\end{example}

\begin{figure}[h!]
\centering
\includegraphics[width=9cm]{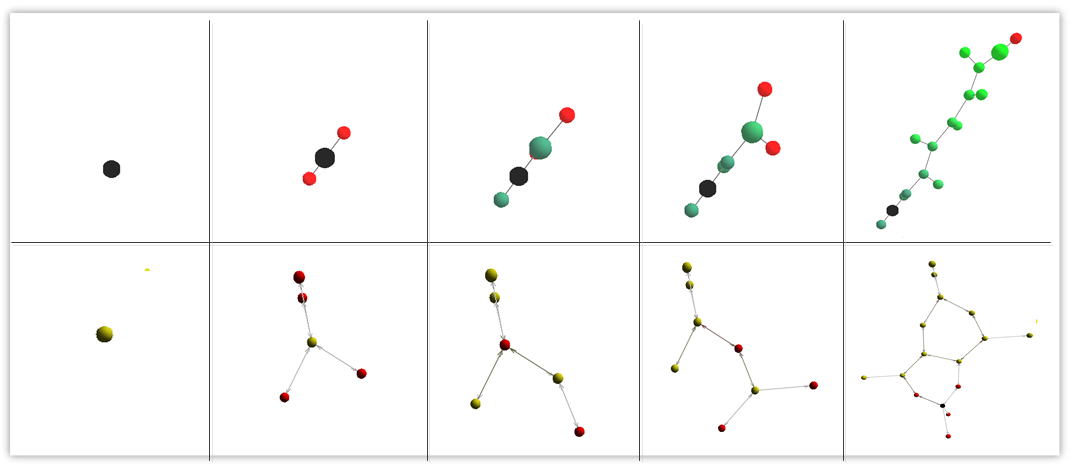}
\caption{Building the proof tree and the Kripke model stepwise}
\label{fig:river_prooftreegraph_step}
\centering
\includegraphics[width=8cm]{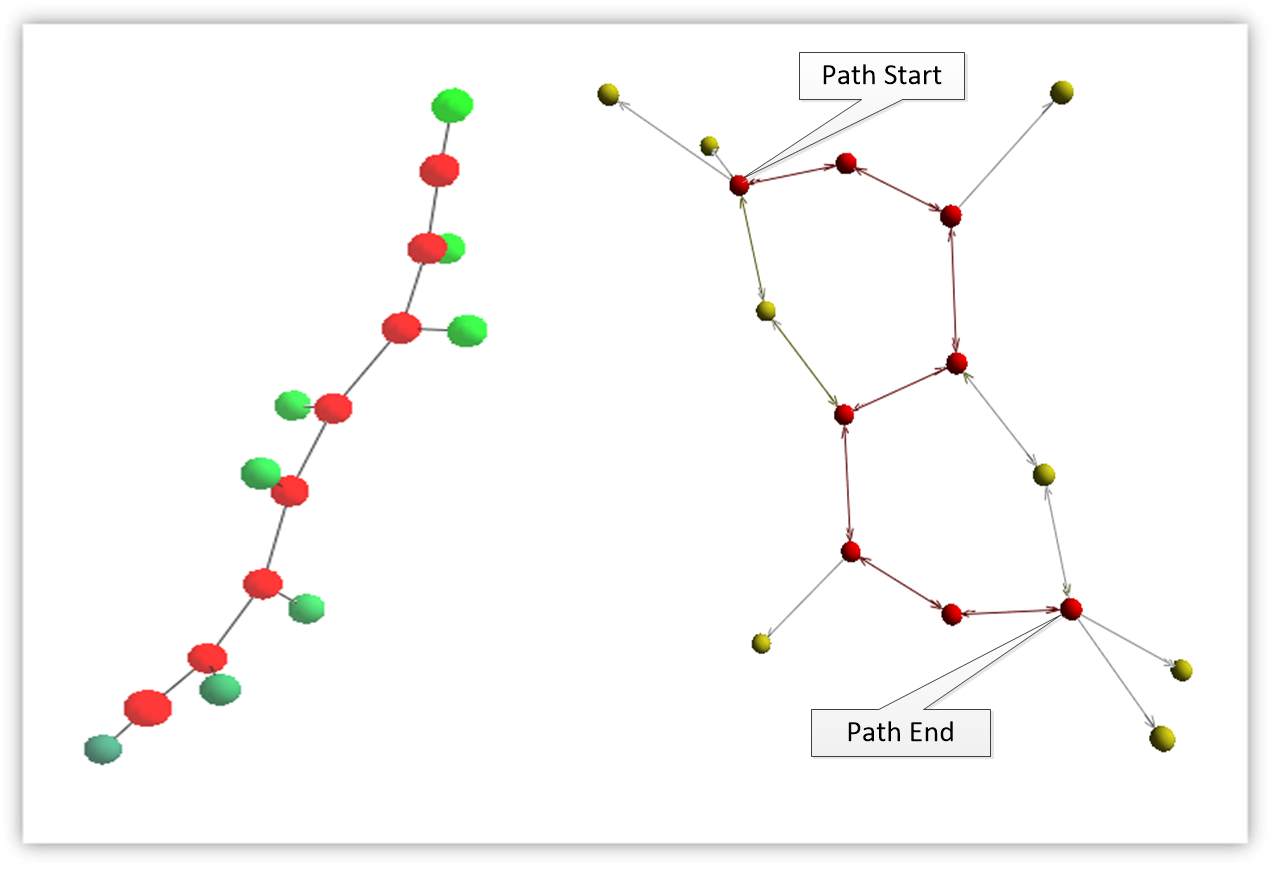}
\caption{The highlighting of all the $EU$-formulas in the proof tree (left), and a path in the model (right).}
\label{fig:river_path}
\end{figure}

We formalize this problem by defining a Kripke model,
where each state is represented by an assignment of four boolean state variables \textsf{farmer}, \textsf{wolf}, \textsf{goat} and \textsf{cabbage}, and
each transition between states is represented by the transformation of an assignment of these variables to another. 
The property to be verified is that whether there is a path that starts with the state 
\begin{center}
$\mathsf{S_0}$: \textsf{\{farmer:false, wolf:false, goat:false, cabbage:false\}}
\end{center}
and end with the state
\begin{center}
$\mathsf{S}$: \textsf{\{farmer:true, wolf:true, goat:true, cabbage:true\}}.
\end{center}
This equals to verify if the sequent $\vdash EU_{x,y}(safe(x), complete(y), S_0)$ is provable in \textsf{SCTL}, taking the Kripke model above as parameter, where $safe(x)$ denotes the atomic formula which specifies that, in state $x$, the sheep is not alone with the wolf, and the cabbage is not alone with the sheep; and $complete(y)$ the atomic formula which specifies that, in state $y$, the farmer has carried all the three items across the river. Using \textsf{VMDV}, we can show in 3D format the proof tree and the Kripke model, provided both by the proof system \textsf{SCTL} (Fig. \ref{fig:river_path}). 
In addition, along with the proof search of the sequent in progress, 
in the Kripke model, a path of states emerges from scratch, certifying this proof.
This is because the formula 
$$EU_{x,y}(safe(x), complete(y), S_0)$$
starts with $EU$ modality (called $EU$-formula), 
and each application of the rule $\mathbf{EU}$-$\mathsf{R_2}$ of \textsf{SCTL} (Fig. \ref{fig:sctl_rules}) 
corresponds to one unfolding step in the Kripke model. This process stops until the $\mathbf{EU}$-$\mathsf{R_1}$ applied. 
During this procedure, \vmdv sends messages to the proof system \textsf{SCTL}, 
and automatically shows the newly constructed nodes both in the proof tree and in the Kripke model (Fig. \ref{fig:river_prooftreegraph_step}).
This way, when we highlight all the nodes with $EU$-formulas in the proof search tree,  
a state path in the Kripke model is also highlighted, certifying this proof (Fig. \ref{fig:river_path}).

\begin{figure}[h!]
\centering
\includegraphics[width=10cm]{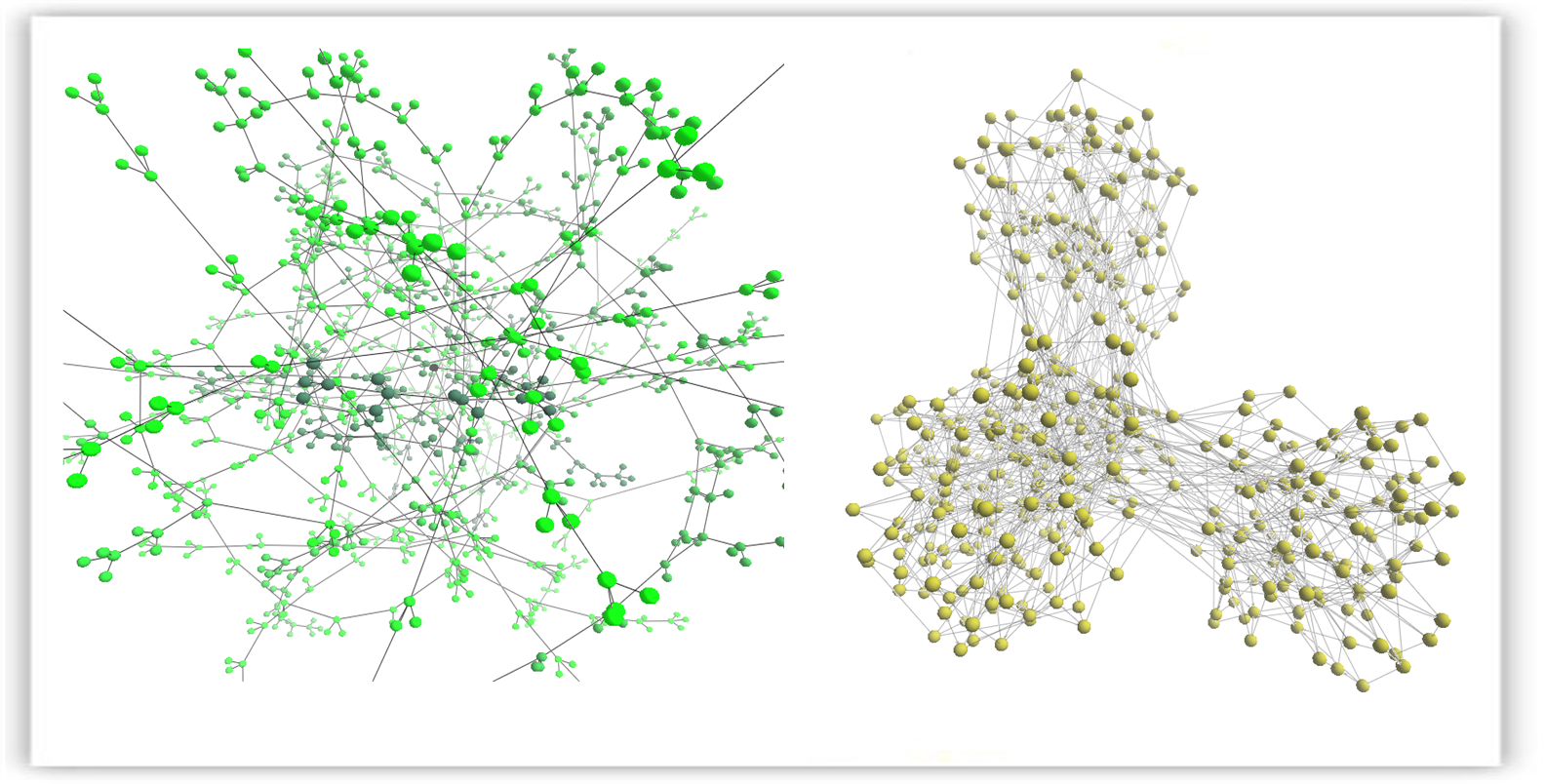}
\caption{The proof tree (left) and the Kripke model (right) for the verification of an $AG$-formula in \textsf{SCTL}.}
\label{fig:ag_proof}
\end{figure}

\begin{figure}[h!]
\centering
\includegraphics[width=12cm]{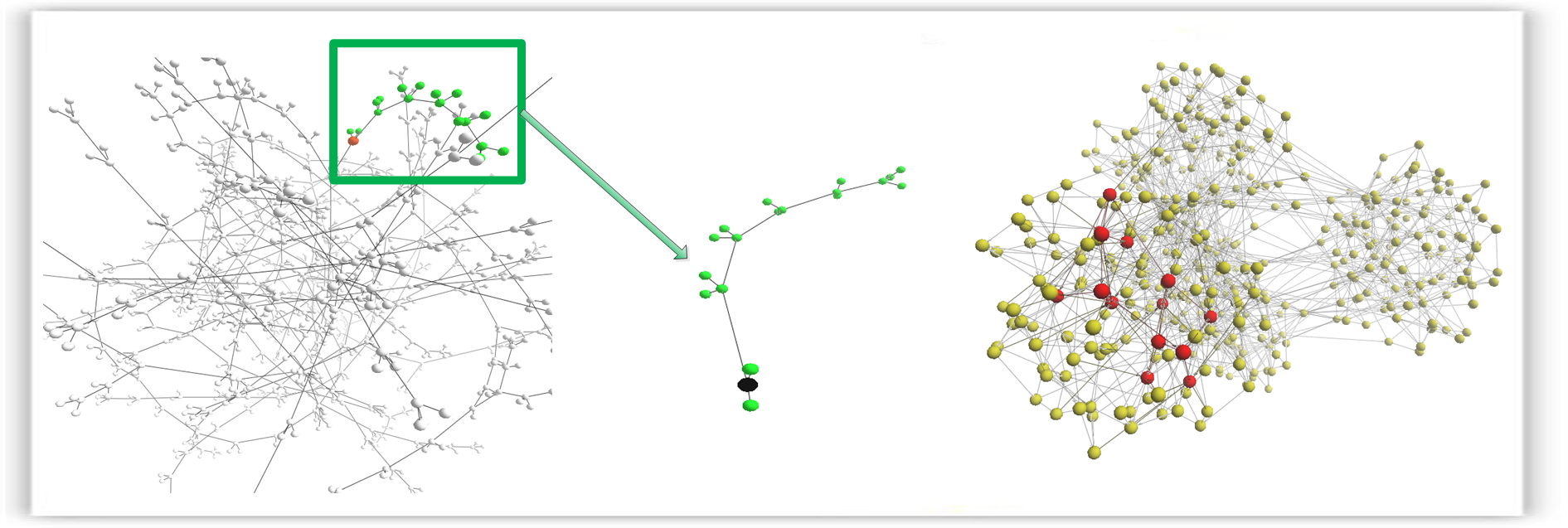}
\caption{Focus on a subtree and highlight related state transitions}
\label{fig:ag_part_detail}
\end{figure}

In \textsf{SCTL}, the proof tree for formulas starting with different modalities are clearly distinguishable from each other,
the same scenario holds for the related part in the Kripke model under consideration. 
For instance, the main part of a proof tree of an $EU$-formula corresponds to a finite path in the model, 
testifying this proof.
While for verifying an $AG$-formula, 
both the proof tree and the related part in the Kripke model may appear very complicated
(Fig. \ref{fig:ag_proof}). 
In this situation, one may focus on a part of the proof tree each time, 
and track the corresponding part of state transitions (Fig. \ref{fig:ag_part_detail}).

\subsection{A Small Example of Coq}
\begin{figure}
\centering
\includegraphics[width=6cm]{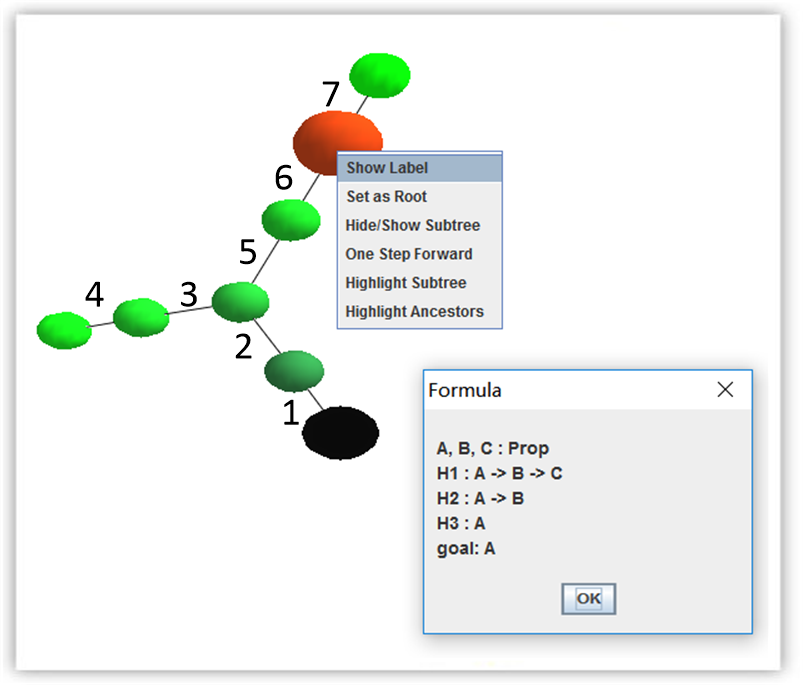}
\caption{Proof tree in the Coq example.}
\label{fig:coq_example}
\end{figure}

We illustrate, by a small example, how \vmdv can visualize proof trees produced by Coq. In Coq, the steps of interactive proof of a formula is controlled by a proof script. Each step of the proof script introduces new proof goals based on the current proof goal. Thus, we formulate the proof tree in such a manner that each proof goal is formulated as a node, and all sub-goals of the current goal are formulated as the sub-nodes of the current node. If the current goal has no sub-goal, then the current node is a leaf node. For instance, the proof script of the formula 
\begin{center}
	$\forall A\; B\; C\; : Prop,\; (A\rightarrow(B\rightarrow C))\rightarrow ((A\rightarrow B)\rightarrow(A\rightarrow C))$
\end{center}
is 
\begin{center}
\begin{verbatim}
Proof.
  intros A B C. intros H1 H2 H3. 
  apply H1. assumption. apply H2. assumption.
Qed.
\end{verbatim}
\end{center}
There are 7 steps (the first ``assumption" comprise two steps: finish the current goal, and jump to the next goal) in this proof script. Thus, the proof tree should have 7 edges, as shown in Fig. \ref{fig:coq_example}.

In order to visualize proof trees produced by Coq in \vmdv, one usually has two options. The first option is to build an integrated development environment (IDE) for the coq toplevel system (coqtop\footnote{\url{https://www.mankier.com/1/coqtop}}) from scratch, and let the IDE communicate with \vmdv. In this situation, the developer may have to handle many basic and cumbersome I/O problems of the IDE and coqtop. The second option is to build a wrapper upon existing interfaces for coqtop, such as coqide\footnote{\url{https://www.mankier.com/1/coqide}}, or Proof General\footnote{\url{https://proofgeneral.github.io/}}. These existing interfaces are usually much more flexible and easy to extend. The developer has to adapt the XML protocol specified by coqide or Proof General and translate the message into the TCP packets that are understandable by \vmdv, which is very straightforward. We prefer the second option, and our interface is now under development. We believe that designing such a interface would guide the interactive proof in Coq in a much more friendly way with visualization, and may be helpful in the education classes. 

\section{Conclusion and Future Works}
In this paper, we have proposed a visualization tool \vmdv for visualising the output of various proof systems in 3D space. 
\couic{
The output from a proof system can be either a proof tree, or some auxiliary structures such as Kripke models 
in the proof system \textsf{SCTL}.
} \vmdv enables us a variety of manners to observe or interact with the 3D graph. In addition, we use an automatic layout algorithm to manage the shape of the 3D graph. 
Up to now, we have made our first step to the application of visualization analysis of proof trees produced by two automatic proof systems. For the future work, we intend to integrate our tool with more sophisticated theorem provers. For instance, in Coq \cite{bertot2013interactive}, the construction of proof trees are controlled by proof scripts, but users can only focus on the current branch of a proof tree, and they have to memorize (or imagine) the overall structure of the current proof tree. 
In this situation, \vmdv can show the overall structure of proof trees along with the dynamically updating of local branches. In this way, users can easily find whether the current formula has been already proved in another branch of the proof tree. If so, there is no need to do extra proof on this formula, and directly abort the proof of the current formula. Designing such a interface for Coq (or other existing proof systems) requires a middleware (broker) that can dispatch control and data messages interchanged by prove engines and \vmdv. This is also the future work.

\newpage
\bibliographystyle{splncs03}
\bibliography{refs}

\begin{thebibliography}{10}
\providecommand{\url}[1]{\texttt{#1}}
\providecommand{\urlprefix}{URL }

\bibitem{CBJPK}
Baier, C., Katoen, J.: Principles of model checking. {MIT} Press (2008)

\bibitem{bajaj2003interactive}
Bajaj, C., Khandelwal, S., Moore, J., Siddavanahalli, V.: {Interactive symbolic
  visualization of semi-automatic theorem proving}. Computer Science
  Department, University of Texas at Austin (2003)

\bibitem{bertot2013interactive}
Bertot, Y., Cast{\'e}ran, P.: {Interactive Theorem Proving and Program
  Development: Coq'Art: The Calculus of Inductive Constructions}. Springer
  Science \& Business Media (2013)

\bibitem{byrnes2009visualizing}
Byrnes, J., Buchanan, M., Ernst, M., Miller, P., Roberts, C., Keller, R.:
  {Visualizing Proof Search for Theorem Prover Development}. Electronic Notes
  in Theoretical Computer Science  226,  23--38 (2009)

\bibitem{CGMZ}
Clarke, E.M., Grumberg, O., McMillan, K.L., Zhao, X.: Efficient generation of
  counterexamples and witnesses in symbolic model checking. In: {DAC}. pp.
  427--432 (1995)

\bibitem{CGP01}
Clarke, E.M., Grumberg, O., Peled, D.: {Model checking}. {MIT} Press,
  Cambridge, MA, USA (2001)

\bibitem{dowek2013logical}
Dowek, G., Jiang, Y.: {A Logical Approach to CTL}  (2013),
  \url{https://who.rocq.inria.fr/Gilles.Dowek/Publi/ctl.pdf}

\bibitem{EmersonC82}
Emerson, E.A., Clarke, E.M.: {Using Branching Time Temporal Logic to Synthesize
  Synchronization Skeletons}. Sci. Comput. Program.  2(3),  241--266 (1982)

\bibitem{EmersonH85}
Emerson, E.A., Halpern, J.Y.: {Decision Procedures and Expressiveness in the
  Temporal Logic of Branching Time}. J. Comput. Syst. Sci.  30(1),  1--24
  (1985)

\bibitem{Farmer200939}
Farmer, W.M., Grigorov, O.G.: Panoptes: An exploration tool for formal proofs.
  Electr. Notes Theor. Comput. Sci.  226,  39--48 (2009)

\bibitem{Fitting96}
Fitting, M.: {First-Order Logic and Automated Theorem Proving, Second Edition}.
  Graduate Texts in Computer Science, Springer (1996)

\bibitem{yifanhu05}
Hu, Y.: Efficient and high quality force-directed graph drawing. The
  Matematical Journal  10,  149--160 (2005)

\bibitem{LibalRR14}
Libal, T., Riener, M., Rukhaia, M.: Advanced proof viewing in prooftool. In:
  Proceedings Eleventh Workshop on User Interfaces for Theorem Provers, {UITP}
  2014, Vienna, Austria, 17th July 2014. pp. 35--47 (2014)

\bibitem{Loveland78}
Loveland, D.W.: {Automated Theorem Proving: A Logical Basis (Fundamental
  Studies in Computer Science)}. Elsevier (1978)

\bibitem{martin2011openord}
Martin, S., Brown, W.M., Klavans, R., Boyack, K.W.: Openord: an open-source
  toolbox for large graph layout. SPIE Proceedings  7868 (2011)

\bibitem{jacomy2014forceatlas2}
Mathieu, J., Tommaso, V., Sebastien, H., Bastian, M.: Forceatlas2, a continuous
  graph layout algorithm for handy network visualization designed for the gephi
  software. PLOS ONE  9(6),  e98679 (2014)

\bibitem{sakurai2011mikibeta}
Sakurai, K., Asai, K.: {MikiBeta: A General GUI Library for Visualizing Proof
  Trees}. In: Logic-Based Program Synthesis and Transformation, pp. 84--98.
  Springer (2011)

\bibitem{steel2005visualising}
Steel, G.: {Visualising First-Order Proof Search}. In: Proceedings of User
  Interfaces for Theorem Provers. vol. 2005, pp. 179--189 (2005)

\end{thebibliography}



\end{document}